# Defect states and disorder in charge transport in semiconductor nanowires

Dongkyun Ko[1], X. W. Zhao[1], Kongara M. Reddy[2], O. D. Restrepo[2], R. Mishra[2], I. S. Beloborodov[3], Nandini Trivedi[1], Nitin P. Padture[2], W. Windl[2], F. Y. Yang[1,†]

and E. Johnston-Halperin[1,†]

[1]*Department of Physics, The Ohio State University, Columbus, Ohio 43210, USA*

[2]*Department of Materials Science and Engineering, The Ohio State University, Columbus, Ohio 43210, USA*

[3]*Department of Physics and Astronomy, California State University at Northridge, Northridge, California 91330, USA*

Abstract

We present a comprehensive investigation into disorder-mediated charge transport in InP nanowires in the statistical doping regime. At zero gate voltage transport is well described by the space charge limited current model and Efros-Shklovskii variable range hopping, but positive gate voltage (electron accumulation) reveals a previously unexplored regime of nanowire charge transport that is not well described by existing theory. The ability to continuously tune between these regimes provides guidance for the extension of existing models and directly informs the design of next-generation nanoscale electronic devices.

PACS numbers: 72.20.-i, 73.20.At, 73.63.Bd, 71.23.-k

Semiconducting nanostructures and nanowires are the focus of extensive efforts to develop next-generation electronic devices, ranging from single-nanowire field-effect transistors (FETs) [1-3] to multi-spectral and multi-platform optoelectronic devices [4]. However, along with the technological promise afforded by their novel properties (for example: high strain tolerance [5], enhanced gate sensitivity [6] and vertical geometry [7]) comes the scientific challenge of extending our knowledge of bulk and thin-film properties to the extreme surface to volume ratios, statistical regime for defects and dopants [8, 9] and enhanced sensitivity to disorder [10] inherent in these quasi-1D structures. Previous work has begun to address this challenge in various regimes of space-charge limited current (SCLC) [11,12] and hopping transport [13,14]; however, a detailed exploration of the validity of existing models, and the experimental foundation for extending that scope, is lacking. Here we present a comprehensive experimental investigation of nanowire transport in the presence of disorder as a function of source-drain bias, gate potential and temperature. Gate-control of the carrier density allows access to regimes that are well described by existing theory (such as the SCLC model [15] and Efros-Shklovskii variable range hopping [16]) and those that are not within the same structure. The ability to tune continuously between these regimes identifies promising directions for the extension of existing models and future experiments, as well as directly informing the design of next-generation nanoscale electronic devices.

Se-doped InP nanowires are grown in a pulsed laser deposition (PLD) system based on previously developed recipes for GaAs nanowire growth (supplemental material 1(a) [34], [17, 18]). Typical nanowire diameter is 50 ~ 60 nm and typical length is >10 μm. Figure 1 shows high resolution transmission electron microscopy (HRTEM) images of a representative InP nanowire. The core is found to have zinc-blende structure with growth along the <111> axis and an outer oxide shell 2 - 5 nm in thickness.

For electrical measurements, the nanowires are removed from the growth substrate by sonication in methanol and are dispersed onto a $SiO_2$/Si wafer (300nm/450μm). Isolated high-quality nanowires are identified using scanning electron microscopy (SEM) and single-wire field effect transistors are fabricated using standard fabrication techniques (supplemental material 1(b) [34]). In Fig. 2, 2-probe and 4-probe current-voltage (I-V) measurements show that the contact resistance is small compared to the nanowire channel resistance (all subsequent data is collected in the 4-probe geometry). Despite apparently Ohmic response, the current as a function of gate voltage, $V_g$, shows saturation at negative gate voltage and *n*-type behavior turning on at a positive gate voltage of approximately +9 V (Fig. 2, bottom right inset). This gate response stands in contrast to optimized nanowire FET structures that exhibit similar linear I-V curves but show pinch-off at negative gate voltage with on/off ratios of $10^4$ - $10^5$ [4], but is consistent with studies of nanowire systems with high trap density [19, 20].

Further evidence for defect-mediated transport can be found in the temperature dependence of the I-V characteristic at zero gate voltage, Fig. 3(a). The nonlinearity is exacerbated at low temperature, and quantitative comparison to a Schottky model [21] and plotting on a semi-log scale (Fig. 3(a), inset) indicate that this nonlinearity is not Schottky in origin. Indeed, the linearity at high bias seen in *log-log* plots (Fig. 3(b)) suggests that a power law, rather than the exponential dependence predicted by the Schottky model, more appropriately describes electronic transport in these systems.

There are two key features of the data in Fig. 3(b) that provide further insight into the transport mechanism: i) the slope is temperature dependent and ii) linear extrapolations of the data for high temperatures (> 150 K) converge to a single, temperature independent, voltage, $V_c$. This behavior suggests the trapped space-charge limited current (SCLC) model [15], wherein injected electrons are trapped by defect states and the amount of trapped charge depends on temperature and source-drain bias ($V_{s-d}$). Expressed in the Arrhenius form to highlight the temperature dependence of the current density the SCLC model predicts [22]:

$$J = \left(\frac{\mu q N V_{s-d}}{2d}\right) \exp\left[-\frac{E_t}{k_B T} \ln\left(\frac{q H_t d^2}{2\varepsilon V_{s-d}}\right)\right] \quad (1)$$

where $\mu$ is the electron mobility, $d$ is the channel length and $N$ is the band density of states. Taking $E_t = k_B T_t$ at the crossover voltage, $V_c$, the activation energy can then be written as:

$$E_a = E_t \ln\left(\frac{q H_t d^2}{2\varepsilon V_c}\right) \quad (2)$$

At $V_c$ the current density is temperature independent, allowing us to solve for $H_t$ given the known nanowire length and permittivity ($H_t = 2.0 \times 10^{16}$ cm$^{-3}$ for $V_c = 20$ V, Fig. 3(b)).

This analysis in turn allows us to better understand the gate dependence in Fig. 2. If we interpret the transition at $V_g = +9$ V as the Fermi energy moving from the defect states into the conduction band of InP, then at $V_g = +9$ V the electron density is equal to the trap density, i.e. $n_e = 2.0 \times 10^{16}$ cm$^{-3}$. Further, the calculated capacitance of the nanowire FET (0.14 fF) allows determination of the carrier concentration at any arbitrary bias. More specifically, the carrier density at zero gate bias is $1.3 \times 10^{16}$ cm$^{-3}$. Given the 2 μm length and 50 nm diameter of the channel this corresponds to $N_e \sim 50$ and $\sqrt{N_e} \sim 7$ (where $N_e$ is the total number of electrons), placing these nanowires solidly within the statistical doping regime [8]. This hypothesis is further supported by the dependence of the SCLC regime on gate voltage. The SCLC model is valid where band carriers and trap states coexist, and this regime of validity extends to lower $V_{s-d}$ and lower temperature as the gate voltage, and consequently $n_e$, increases above $V_g = +9$ V (the opposite trend is seen for gate bias decreasing below $V_g = +9$ V; see Fig. S1 in supplemental material 2 [34]).

This model implicitly assumes the presence of relatively shallow trap states, i.e. states within a few times $k_BT$ of the conduction band with coupling to the band states. In order to further assess the plausibility of this assumption, we determine the formation energy of competitive defects in "bulk" InP from density-functional theory (DFT; calculation of the interface/surface states would require detailed atomic-scale structural characterization for a realistic picture and is not considered here). The calculations are performed for supercells with 108 formula units using the VASP package and potentials [23]; the Generalized-Gradient Approximation (GGA) [24], cross-checked by the HSE06 hybrid-functional [25]; and a 2×2×2 $k$-points mesh for the Brillouin zone integration. Charge states are treated in analogy to Ref. [26]. XPS measurements suggest that the native oxide on InP is P rich at the interface, and that the oxidation reaction happens at the surface [27]. Thus, we assume In-rich conditions in the nanowire for determining the chemical potentials for defect calculations. Using the method of constitutional defects [28], we find $In_P^{-1}$ and $O_P^0$ to be the most stable defects that also create a trap state near the conduction band edge. Assuming that the oxidation reaction indeed happens at the surface [27], it is however less likely that $O_P^0$ defects will form (and if they do, are found to form an atomic-like deep level), leaving $In_P^{-1}$ with a shallow level ~60 meV below the conduction-band edge (see Fig. 3(b), inset) as a plausible candidate for the trap states identified above [29].

Next, we consider the regime where the SCLC model is *not* valid, i.e. low $V_{s-d}$ and low temperature where the Fermi energy lies in the defect states and hopping transport is expected. Generally accepted models of hopping transport follow the form [16]:

$$R = R_0 \exp(T_0/T)^m \qquad (3)$$

where $R_0$ indicates the overall resistance of the channel, $T_0$ relates to the energy scale of the hopping transport and $m$ is a constant less than or equal to 1 that indicates the mechanism of the hopping transport. In order to extract the values of these constants for the InP nanowires studied here we plot the zero-bias resistance on a *log(ln(R))* vs. *log(1/T)* scale for various gate voltages (Fig. 4(a)). At zero gate voltage (red-squares) the data reveal two slopes, $m_{high}$ = 1.03 ± 0.02 at T > 158 K and $m_{low}$ = 0.49 ± 0.02 at T < 158 K. The crossover temperature (158 K at $V_g$ = 0 V) depends on gate voltage and shifts to lower temperature with increasing positive gate voltage. For negative gate voltage, the shape of the curve becomes more complex and does not exhibit a well-defined crossover.

The slopes at $V_g$ = 0 V indicate nearest-neighbor hopping (NNH) and Efros-Shklovskii variable range hopping (ES-VRH), respectively [30, 31]. The inset in Fig. 4(a) schematically depicts the crossover between these regimes. At high temperature ($k_BT$ greater than the average variation in trap energy) phonon-assisted hopping allows access to NN hopping sites, as the temperature decreases the range of energetically accessible hopping sites decreases and VRH dominates. The energy scale, $T_{NNH}$, at high temperature is found to be 58 meV (in good agreement with our DFT calculations of the $In_P^{-1}$ defect), while the energy scale at low temperature is estimated as $T_{ES\text{-}VRH}$ ~ 230 meV (this corresponds to a hopping energy that varies from 32 meV at 300 K to 19 meV at 100 K). Finally, using the standard relation for the localization length, $\xi = \beta e^2 / k_B \varepsilon T_{ES-VRH}$, for ES-VRH [32] we determine that in this regime $\xi$ ~ 23 nm for $\beta = 3.9$ and $\varepsilon = 12\varepsilon_0$ ($\beta$ is determined by setting the VRH and NNH distance equal at $T = 158$ K and the dielectric constant is the bulk value for InP).

When the crossover temperature between NNH and ES-VRH, $T_{cr}$, is plotted as a function of gate voltage the same critical value of +9 V is identified as a point of departure from the physics at zero gate voltage (Fig. 4(b)). This trend is confirmed when similar analysis is performed for $m_{low}$, $T_{NNH}$ and $T_{ES\text{-}VRH}$. While this variation lies beyond the scope of current theory, we can gain a qualitative understanding by referring to the two-channel model presented above. At high temperature NNH, or Arrhenius, transport is characterized by thermal activation to the conduction band followed by re-trapping at a NN site. In this regime the increased gate voltage merely increases the number of carriers capable of executing these hops, resulting in a relatively weak perturbation from the zero voltage case ($T_{NNH}$ increases by about 12% and $m_{high}$ does not vary to within experimental error).

In contrast, at low temperature the hopping transport undergoes a more dramatic change as the Fermi energy enters the conduction band (the increase in $T_{ES\text{-}VRH}$ is ~300%). Focusing on the behavior of $m_{low}$, one might expect an *increase* towards $m = 1$ as the presence of carriers in the conduction band opens a channel for NNH or band transport; however, what is actually observed is a *decrease* in $m_{low}$. In attempting to understand this result, it is useful to consider two structural shortcomings of both the NNH and ES-VRH models. First, both models assume that the transport physics can be described by a single conduction channel with well-defined energy scale, localization length, etc. However, our analysis of the SCLC regime indicates that for $V_g > +9$ V both the trap states and the conduction band are occupied. Moreover, while transport through the defect states proceeds through ES-VRH (as demonstrated by the $V_g = 0$ V data in Fig. 4(a)), transport within the conduction band occurs in the presence of a disorder potential defined by the spatial distribution and occupancy of the trap states. This disorder potential can lead to the breakup of the conduction band into grains (supplemental material 3 [34], [33]) and at low temperature and bias likely leads to hopping transport in the conduction band as well. As a result, a full theoretical description of the transport in this regime needs to account for two parallel conduction channels that are correlated by the interplay between the trap states and the disorder potential in the conduction band.

Second, and more generally, the low-temperature low-$V_{s-d}$ regime is characterized by a competition between correlation effects and disorder. The average separation between carriers, $r_0$, is comparable to the thermal deBroglie wavelength for all temperatures considered here (roughly 25 nm and 16 – 27 nm, respectively; supplemental material 4 [34]). As a result, this system is well into the quantum regime while both NNH and ES-VRH treat electron-electron interactions classically. The primary effect of these quantum correlations will be to modify the low-energy density of states, suggesting future studies exploiting tunneling physics to directly probe this variation.

In conclusion, our comprehensive exploration of defect- and disorder-mediated transport in semiconducting nanowires reveals a rich interplay between localized trap states, conduction band transport, quantum effects and electron-electron correlations in the statistical doping regime. These experimental results provide a foundation for extending current models of hopping transport to account for multiple conduction channels and the competition between correlation and disorder as well as suggesting promising directions for further experiments to determine the detailed nature of the low-energy density of states. In turn, these fundamental studies provide critical insight into the transport mechanisms likely to be found in next-generation nanoscale electronic devices.

We would like to acknowledge Mr. Kurtis Wickey for critical reading of the manuscript and fruitful discussions concerning hopping models of electron transport. Support for this research is provided by the Center for Emergent Materials at The Ohio State University, an NSF MRSEC (DMR-0820414), DOE (DE-SC0001304), the IMR at The Ohio State University and the Ohio Supercomputer Center for computer time.

**References**

[†]To whom correspondence should be addressed. E-mail: fyyang@mps.ohio-state.edu and ejh@mps.ohio-state.edu.

[1] A. Appenzeller *et al*., IEEE Trans. Electron Devices **55**, 2827 (2008).

[2] W. Lu and C. M. Lieber, Nature Mater. **6**, 841 (2007).

[3] C. Thelander et al., Nano Lett. **5**, 635-638 (2005).

[4] X. Duan *et al*., Nature **409**, 66 (2001).

[5] E. Lind *et al*., Nano Lett. **6**, 1842 (2006).

[6] S. N. Cha *et al*., Appl. Phys. Lett. **89**, 263102 (2006).

[7] T. Bryllert *et al*., IEEE Electron Device Lett. **27**, 323 (2006).

[8] S. Markov *et al.*, IEEE Trans. Electron Devices **57**, 3106 (2010).

[9] S. Zhang *et al*., Nano Lett. **9**, 3268 (2009).

[10] G. Micocci *et al*., J. Appl. Phys. **82**, 2365 (1997).

[11] A. D. Schricker *et al*., Nanotechnology **17**, 2681 (2006).

[12] Y. Gu and L. J. Lauhon, Appl. Phys. Lett. **89**, 143102 (2006).

[13] Y. J. Ma *et al*., Nanotechnology **16**, 746 (2005).

[14] M. M. Fogler *et al*., Phys. Rev. B **69**, 035413 (2004).

[15] P. Mark and W. Helfrich, J. Appl. Phys. **33**, 205 (1962).

[16] M. Pollak and B. I. Shklovskii, *Hopping Transport in Solids* (North-Holland, Amsterdam, 1991).

[17] X. W. Zhao *et al*., Nanotechnology **18**, 485608 (2007).

[18] L. Fang *et al*., In submission.

[19] S. A. Dayeh *et al*., J. Vac. Sci. Technol. B **25**, 1432 (2007).

[20] J. Maeng *et al*, Mat. Res. Bull. **43**, 1649-1656 (2008).

[21] S. M. Sze, *Physics of semiconductor devices* (John Wiley & Sons, New York, 1981).

[22] V. Kumar *et al*., J. Appl. Phys. **94**, 1283 (2003).

[23] G. Kresse and J. Furthmüller, Phys. Rev. B 54, 11169 (1996); P.E. Blöchl, Phys. Rev. B **50**, 17953 (1994); G. Kresse, and D. Joubert, Phys. Rev. B **59**, 1758 (1999).

[24] J. P. Perdew and Y. Wang, Phys. Rev. B **45**, 13244 (1992).

[25] J. Heyd *et al*., J. Chem. Phys. **118**, 8207 (2003); **124**, 219906 (2006); J. Paier *et al*. **124**, 154709 (2006).

[26] W. Windl *et al*., Phys. Rev. Lett. **83**, 4345 (1999).

[27] G. Hollinger *et al*., J. Vac. Sci. Technol. B **5**, 1108 (1987).

[28] R. Mishra *et al*. Chem. Mater. 22, 6092 (2010).

[29] O. D. Restrepo *et al*., to be published.

[30] A. Yildiz *et al*., Jpn. J. Appl. Phys. **48**, 111203 (2009).

[31] A. L. Efros and B. I. Shklovskii, J. Phys. C **8**, L49 (1975).

[32] R. Rosenbaum, Phys. Rev. B **44**. 3599 (1991).

[33] I. S. Beloborodov *et al*., Rev. Mod. Phys. **79**, 469 (2007).

[34] See supplemental material at --------- for further experimental conditions and quantitative analysis.

FIG. 1. (a) and (b) are high resolution TEM images revealing single crystal structure and a thin oxide layer (2~5nm) on the surface of the nanowire. (c) Diffraction pattern verifying the zinc-blende structure of InP nanowire.

FIG. 2. (color online). 2-probe and 4-probe I-V measurements show that contact resistances are less than 0.05 M$\Omega$. Left-top inset: SEM image of single nanowire field effect transistor (FET) device with 4 electrodes. Right bottom inset: $I$ vs. $V_g$ showing saturation current at negative gate voltage.

FIG. 3. (color online). (a) Temperature dependent I-V plots and fitting to Schottky model (ideality factor ranges from 97 to 73). Inset is the semi-log plot of showing non-exponential function dependence on $V_{s-d}$. (b) A log-log plot of the same data shows linear behavior, $I \sim V^S$, with slope increasing as temperature decreases. The extrapolations of the linear fits converge to a crossover point ($V_c$, see text). Inset is a band-structure plot of the $In_P^{-1}$ defect using GGA.

FIG. 4. (color online). (a) Log-log plot of $ln(R)$ vs. $1/T$. The graph shows that there is a crossover in slope: from $m_{high}$ = 1.03 at high temperature to $m_{low}$ = 0.49 at low temperature ($V_g$ = 0 V). Inset is a cartoon showing nearest-neighbor hopping (NNH) at high temperature and ES variable range hopping (ES-VRH) at low temperature (see text). (b) Red dot is the low temperature slope deviating from $m$ = 0.5 at $V_g$ > +9 V. Similar gate voltage dependent trends

can be seen in the crossover temperature ($T_{cr}$), NNH temperature ($T_{NNH}$) and ES-VRH temperature ($T_{ES\text{-}VRH}$) vs. $V_g$ plots.

FIG.1.

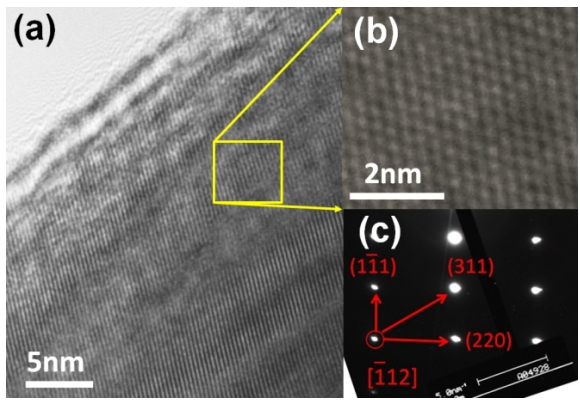

FIG. 2.

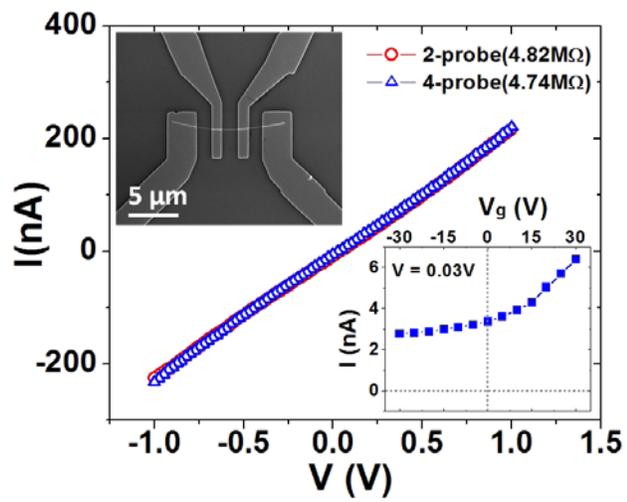

FIG. 3.

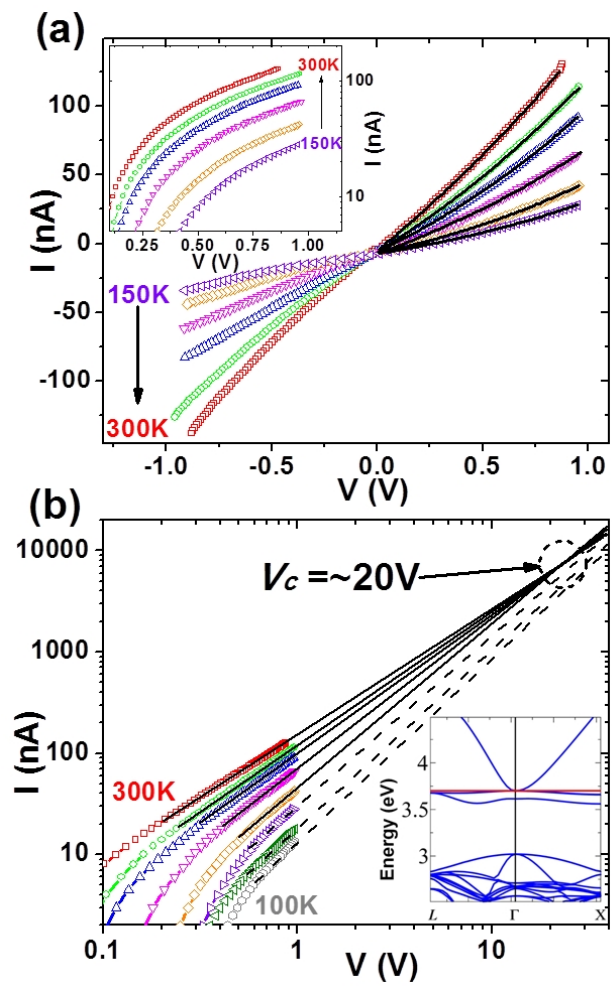

FIG. 4.

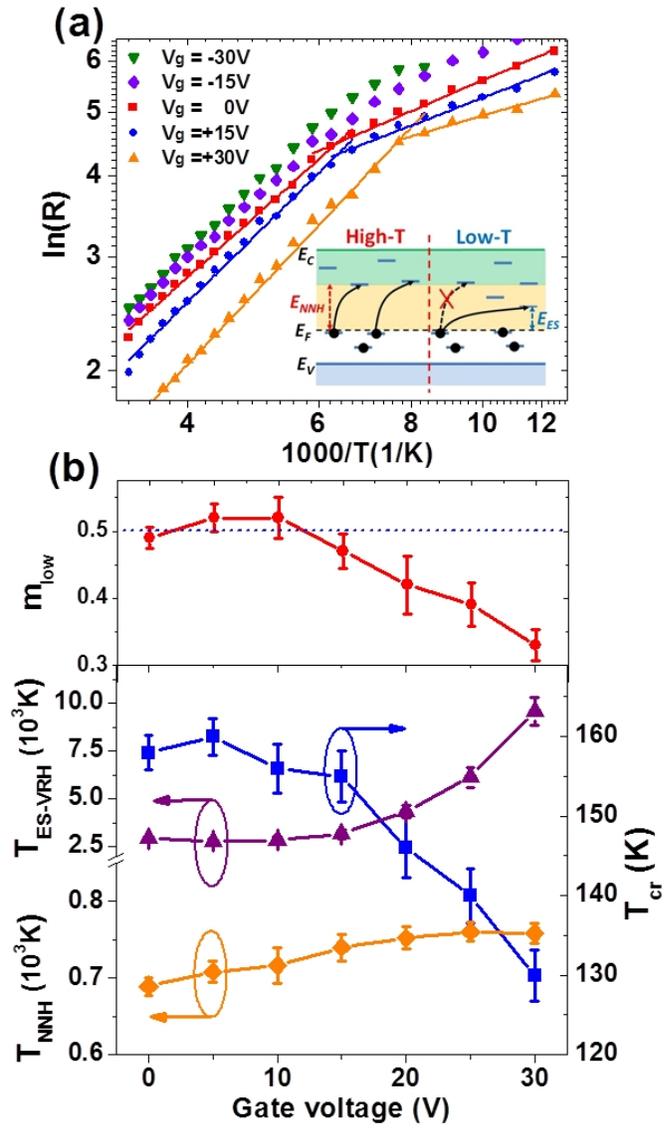

# Supplemental material for

# Defect states and disorder in charge transport in semiconductor nanowires

Dongkyun Ko[1], X. W. Zhao[1], Kongara M. Reddy[2], Oscar D. Restrepo[2], Rohan Mishra[2], I. S. Beloborodov[3], Nandini Trivedi[1], Nitin P. Padture[2], W. Windl[2], F. Y. Yang[1] and E. Johnston-Halperin[1]

[1]*Department of Physics, The Ohio State University, Columbus, Ohio 43210, USA*

[2]*Department of Materials Science and Engineering, The Ohio State University, Columbus, Ohio 43210, USA*

[3]*Department of Physics and Astronomy, California State University at Northridge, Northridge, California 91330, USA*

## Supplemental material 1: Nanowire growth and sample fabrication

### a. Nanowire growth

50 nm gold colloid is dispersed onto a silicon substrate as a catalyst for vapor-liquid-solid (VLS) growth. A 1% Se/InP target is prepared by mixing and pressing InP and $In_2Se_3$ polycrystalline powder. The growth temperature of the substrate is 480ºC and the pressure is controlled at 50~100 Torr with a flow rate of 50 sccm with argon as the carrier gas. A 2 Hz pulsed excimer laser with a wavelength of 248 nm and duration of 10 ns is focused on the target to trigger ablation upstream of the substrate and produces In, P and Se atomic vapors that are subsequently swept across the growth region. Typical nanowire diameter is 50 ~ 60 nm and length is > 10 μm.

### b. Sample fabrication

For electrical measurements, the nanowires are removed from the growth substrate by sonication in methanol and are dispersed onto a $SiO_2$/Si wafer (300nm/450μm). Isolated high-quality nanowires are identified using scanning electron microscopy (SEM) and marked using platinum alignment markers deposited using *in situ* focused ion beam (FIB) decomposition of an organometallic precursor ($C_9H_{16}Pt$). Electrical contacts are defined by briefly removing the sample from the SEM, spin coating poly-methyl methacrylate (PMMA; 4% in anisol), and reloading into the SEM for electron-beam lithography indexed to the platinum alignment markers. Metalization consists of an Ohmic stack, Ge/Au/Ni/Au (2nm/20nm/50nm/50nm), directly after an HCl dip to etch the native oxide shell. Finally, rapid thermal annealing (RTA) in a forming gas environment is done to reduce contact resistance between the metal stack and the nanowire.

# Supplemental material 2: Gate dependence of SCLC behaviors

Figs. S1a – S1c show the I-V characteristics of a nanowire FET on a log-log scale for *positive* gate voltages. The region of linearity extends to increasingly lower bias and lower temperature as gate voltage increases and the region of validity for the SCLC model increases (coexistence of band carriers and traps, see main text). Figs. S1e – S1f show similar plots for *negative* gate voltages (all plots have the same current and voltage scale). The region of linearity for these gate voltage shows opposite behavior trend (shrinking as gate voltage increases in the negative direction) and pure hopping transport dominates SCLC.

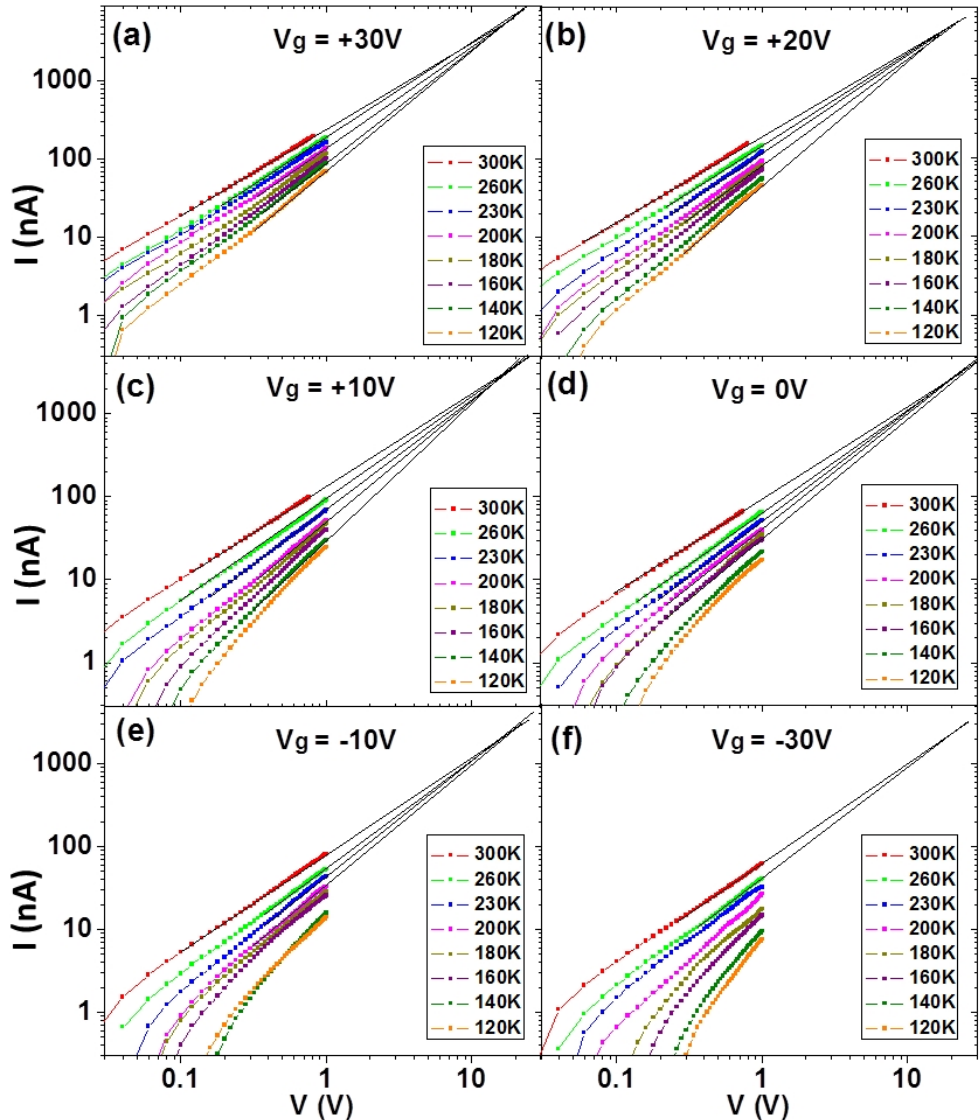

**Fig. S1**: (a) – (f) I-V characteristics on log-log scale for different gate voltages. At the largest negative gate voltage (-30 V, panel (f)) the Fermi energy lies deep in the trap states, far from the conduction band, and hopping transport dominates at all but the highest temperatures and highest source-drain bias. As the gate voltage increases towards positive values mixed band- and trap-mediated transport (the SCLC regime) dominates to increasingly lower temperatures and lower source-drain bias. At a gate voltage of +30 V (panel (a)) the SCLC regime dominates almost the entire measurement window. (Panel (d) is the same data as shown in the main text, Fig. 3b)

## Supplemental material 3: Absence of Mott-VRH and disorder potential at conduction band

We expect in highly disordered nanowires Coulomb interactions should dominate at low temperatures. Mott variable range hopping arises from non-interacting electrons in a random potential gives an exponent $m = 1/(d+1)$ and equals ½ in d=1, whereas ES-VRH that arises in the classical limit from the combined role of disorder and Coulomb interactions also gives an exponent of ½ independent of dimensionality. However we believe that the physics of these low dimensional wires is determined by the combined effects of correlations and disorder (see Supplemental Material 4) and therefore expect that ES-VRH, rather than Mott VRH, dominates in our samples.

In addition, it is possible that the random distribution of charged traps give rise to a granular morphology in the conduction band (see main text). Indeed, in granular samples the role of the Coulomb interaction is strongly enhanced and thus Mott VRH is difficult to observe. This can be understood as follows: in semiconductors, the Efros-Shklovskii law may turn to the Mott behavior with the increase of temperature. This happens when the typical electron energy $\varepsilon$ involved in a hopping process becomes larger than the width of the Coulomb gap $\Delta_c$, i.e., when it falls into the flat region of the density of states where Mott behavior is expected. To estimate the width of the Coulomb gap $\Delta_c$, one compares the ES expression for the density of states

$$\nu(\Delta_c) \propto (\kappa/e^2)^d |\Delta_c|^{d-1}, \qquad (1)$$

with the bare density of states $\nu_0$ i.e., the DOS in the absences of the long-range part of the Coulomb interactions. Using the condition $\nu(\Delta_c) \propto \nu_0$ we obtain

$$\Delta_c = \left(\frac{\nu_0 e^{2d}}{\kappa^d}\right)^{1/(d-1)}. \qquad (2)$$

Inserting the value for the bare DOS, $\nu_0 = 1/E_c \xi^d$ ($E_c$ is the charging energy for a single grain), into Eq. (2) we finally obtain

$$\Delta_c \propto E_c. \qquad (3)$$

Equation (3) means that there is no flat region in the density of ground states and, thus, the Mott regime is difficult to observe in granular wires. To conclude this section we present some estimates for the Coulomb gap $\Delta_c$ and the charging energy $E_c$. The typical grain sizes in our nanowires are in the range 5 nm < a < 20 nm. These grain sizes are justified by the facts that 1) our samples are stable meaning that each nanowire has more than one grain in diameter, and 2) our data clearly show the variable range hopping behavior; this behavior may not hold for a nanowire with a single grain in diameter. Using these numbers for the charging energies of a single grain $E_c = e^2/\kappa a$ we obtain $\frac{10^2}{\kappa} K < E_c < \frac{10^3}{\kappa} K$. The typical dielectric constant $\kappa$ for our samples is 3-4 reflecting the fact that our samples are pure conductors. We would like to point out that the charging energy $E_c$ is larger than the characteristic energy $e^2/\kappa r_{hop}$ scale related to the typical electron hop $r_{hop}$. This is a consequence of the fact that the typical hoping distance $r_{hop}$ is several times larger than the characteristic size of a single grain $r_{hop} > a$. Physically this inequality means that an electron propagates through several grains in one hop.

## Supplemental material 4: Calculation of $r_0$ and $\lambda_t$

The average separation between carriers ($r_0$) at zero gate voltage can be calculated with the simple relation:

$$\frac{4}{3}\pi (r_0)^3 = \frac{1}{n_e}$$

For a typical sample at $V_g = 0$ V we have $n_e = 1.3 \times 10^{16}$ cm$^{-3}$ (see main text), giving $r_0 \sim 25$ nm.

At the same time, the thermal deBroglie wave length is given by:

$$\lambda_t = h \Big/ \sqrt{2\pi m_e k_B T},$$

where $h$ is Plank constant, $m_e$ is the effective mass of electron in InP ($m_e = 0.08 \times m_0$) and $k_B$ is Boltzmann's constant.

In our samples $\lambda_t$ varies from 27 nm at T = 100 K to 16 nm at T = 300 K, revealing that these two length scales are comparable even at zero gate potential and the system becomes more quantum mechanical with increasing gate voltage ($r_0 \sim 22$ nm at $V_g = 9$ V). By this simple estimate, our samples exist in the quantum regime for the entire phase spa